\documentclass[submission,copyright]{eptcs}
\providecommand{\event}{F-IDE 2014} 
\usepackage{graphicx}
\usepackage{enumitem}
\usepackage{wrapfig}

\title{OpenJML: Software verification for Java 7 using JML, OpenJDK, and Eclipse}
\author{David R. Cok
\institute{GrammaTech, Inc.\\
Ithaca, NY, USA}
\email{cok@frontiernet.net}
}
\def\titlerunning{OpenJML}
\def\authorrunning{David R. Cok}
\begin{document}
\maketitle

\begin{abstract}
OpenJML is a tool for checking code and specifications of Java programs.
We describe our experience building the tool on the foundation
of JML, OpenJDK and Eclipse, as well as on many advances in specification-based software verification. The implementation demonstrates the
value of integrating specification tools directly in the software development IDE
and in automating as many tasks as possible. The tool, though still in progress, has now been used for several 
college-level courses on software specification and verification and for small-scale studies on existing Java programs. 
\end{abstract}

\section{Introduction to OpenJML}

 OpenJML is a tool for verification (checking consistency of code and specifications) of Java programs. It is an implementation of the Java 
 Modeling Language (JML) \cite{Leavens-Baker-Ruby99b,Leavens-etal08} and is built using the OpenJDK \cite{OPENJDKweb} compiler. 
 OpenJML can be used both as a command-line tool and through a GUI built on Eclipse.
 
 The tool has been under development in bursts since 2009. Initially the project was simply an experiment to see whether OpenJDK would make a good replacement for the custom parser that underlies ESC/Java(2) \cite{Leino-Nelson-Saxe00,Kiniry-Cok05-short} and the MultiJava compiler that underlies the JML2 tools, neither of which implemented 
 Java generics. The development was reinvigorated in 2011 and is now supporting academic and trial users and staying current with developments in Java.

OpenJML's long-term goal is to provide an IDE for managing program specifications that naturally fits into the practice of daily software development and so becomes a part of expected
software engineering practice. To that end, the project has several nearer-term objectives, discussed in more detail in the body of this paper:
 \begin{itemize}[noitemsep,nolistsep]
 \item To create an implementation of JML, replacing ESC/Java2;
 \item To provide a tool that can be used in courses and tutorials on software specification and verification;
 \item To provide an integration of many techniques proposed in the software verification research or implemented as standalone proofs of concept over the past few decades;
 \item To be a platform for experimentation:
 \begin{itemize}[noitemsep,nolistsep]
 \item with specification language features
 \item with specification encodings
 \item with SMT solvers on software verification problems
 \end{itemize}
 \item To enable case studies of verification of industrial software;
 \item Eventually, to be a tool to be used in daily, industrial software development. 
 \end{itemize}
OpenJML includes operations for a number of related tasks:
 \begin{itemize}[noitemsep,nolistsep]
 \item parsing and typechecking of JML in conjunction with the corresponding Java code
 \item static checking of code and specifications. OpenJML translates Java + JML specifications into verification conditions that are then checked by SMT solvers.
 \item runtime assertion checking by compiling specifications as assertions into the usual Java .class files. OpenJML uses the OpenJDK compiler, enhancing the processing of source files to add appropriate checks that assertions
 and other specifications hold during execution of the program.
 \item integration with both Eclipse and command-line tools
 \item programmatic access through an API to the internal ASTs, type information, compilation and checking commands, and the results of verification attempts, including counterexamples
 \item (planned): javadoc tool that includes JML documentation
 \item (planned): automatic, high-coverage test generation, integrated with unit testing
 \item (planned): specification discovery (a.k.a.\ invariant inference, function summarization) for Java
 \end{itemize}

This is a long-term project, and so intermediate milestones and accomplishments are important. One such accomplishment was becoming current with Java 7. The accomplishment highlighted in this paper is
that OpenJML has now been successfully used in several academic courses. The important points that are relevant to other tools are the successful use of OpenJDK as a foundation, the integration with Eclipse, and the importance of appropriate UI features in making interaction with formal methods readily accessible to users, in particular, effective display of counterexample information.
 
\section{Design of OpenJML}

OpenJML follows a design adapted from ESC/Java2 and commonly used for software verification systems
\cite{Kiniry-Cok05-short,Leino05,FlanaganSaxe01,Barnett-Leino05-short}).
\begin{itemize}[noitemsep,nolistsep]
\item The procedure specifications are translated into assumptions and assertions interleaved with the Java code, based on the semantics of the specification language. For example, preconditions become assumptions at the beginning of the procedure and postconditions are assertions at the end.
\item The code, assumptions, and assertions are translated into a basic block form that uses 
single-assignment labeling of variables.
\item The basic blocks are translated into compact verification conditions (VCs).
\item The verification conditions are expressed in SMTLIBv2 format.
\item An SMT solver of choice (we used primarily CVC4 \cite{webcvc4}, and demonstrated interoperability with Z3 \cite{DeMoura:2008:ZES:1792734.1792766}) is applied to the VC.
\item If the VC is invalid, a counterexample is obtained from the SMT tool.
\item The logical variables of the counterexample are translated back to source code variables and 
text locations; logical variable values are expressed in programming language terms (cf. section \ref{Eclipse}).
\item The counterexample values and the static ``execution'' path are displayed in the source code editor by hover information and highlighting.
\end{itemize}

OpenJML extends this basic model to also check the generated VC for vacuity or for multiple falsified assertions.
OpenJML constructs a single verification condition (VC) for a method. The default behavior is to check the validity of the entire VC. It is also possible, as some tools do, to check the validity of each distinct path, or the set of paths to each distinct assertion, or other sub-expressions of the full VC. If the full VC turns out to be valid, it may be the desired case of having consistent code and specifications. However, it may also be that some combination of user-supplied specifications creates infeasible paths to particular assertions or to the method exit. Thus, after a determination that the VC is valid, OpenJML
checks that there are feasible paths to the procedure exit and to each assertion. 

If the full VC is invalid then there is some specification assertion that is falsified, but it may not be the only one. The user has the option of requesting further checks. OpenJML computes a path
condition leading to the falsified assertion, appends the negation of the path condition to the
VC, and rechecks the VC. The negation of the path condition prevents the same assertion from being
falsified again (through the same path, or alternately at all), but allows finding other assertion violations later in the path, if they are not totally blocked by the first failed assertion. By repeating this procedure until the accumulated VC is found
valid, all falsifiable assertions in the method under scrutiny are found.

\section{Building on OpenJDK}

OpenJDK \cite{OPENJDKweb} is the result of Sun Microsystem's project to release the Java Development Environment as free, open-source code. The software was first available in 2007 and is now managed by Oracle. OpenJDK is the basis for many projects, tools, and experiments with language features.

The OpenJDK compiler architecture has lent itself well to extension, though at present it is not designed for that purpose. The architecture consists of applying a separate operation, embodied in a Java class, for each of a sequence of compiler phases, each of which is replaceable.
OpenJML registers replacement tools for many of the compiler phases; the replacement tools extend (with Java inheritance) the original tool, adding additional functionality. The OpenJDK phases and OpenJML replacements are shown in Table \ref{Table:phases}.
\begin{table}
\begin{tabular}{|p{.3\textwidth}|p{.64\textwidth}|}
\hline
OpenJDK phase & OpenJML replacement \\
\hline
parsing & augmented to parse JML specifications as well, either in .java files or auxiliary .jml files \\ \hline
entering type symbols into the symbol table & augmented to add model types as well \\ \hline
entering class members into the symbol table & augmented to add model fields and methods as well \\ \hline
annotation processing & annotation processing is not yet used or modified in OpenJML \\ \hline
name/type attribution and type checking & augmented to typecheck JML features \\ \hline
flow checks & augmented to add flow checks on JML constructs \\ \hline
 & OpenJML typecheck tool ends here \\ \hline
 & OpenJML static checking is performed by converting the type-attributed AST to VCs and checking them; static checking does not continue with later phases \\ \hline
 & OpenJML runtime checking inserts a phase here to modify the OpenJDK AST, inserting code to perform the runtime checks, and then continues with the compilation phases \\ \hline
desugaring & [no change] \\ \hline
code generation & [no change] \\ \hline
\end{tabular}
\caption{Compiler phases in OpenJDK and OpenJML}
\label{Table:phases}
\end{table}

The extension of OpenJDK to OpenJML is not quite a pure extension; some modifications of OpenJDK are required. It has required using non-public APIs, which may change but have been stable so far. Quite a few changes of visibility (e.g., private visibility to protected visibility, to enable inheritance) were required. Some refactorings were needed to encapsulate a portion of the functionality within a large method, in order to override just that portion. However,
out of 942 files in the source code portion of OpenJDK langtools, 6 needed bug fixes, 5 needed only visibility changes, 21 needed minor refactoring and visibility changes, 6 needed very moderate changes
in structure or functionality, and only 1 needed significant refactoring and new functionality.

Keeping OpenJDK releases up to date through vendor branches encountered few problems. Transition to Java 8 will likely be more difficult and has yet to be attempted.

The main disadvantage of using OpenJDK in conjunction with an IDE is that it is designed as an execute-once compiler. This works efficiently for command-line operations, but is less appropriate for interactive operations through an IDE.
An IDE for software verification should track dependencies, so that the tool automatically knows which modules should be re-verified when a given change is made to the software. OpenJML currently does not have that ability, though it is on the list for future work.

\section{Building on Eclipse}
\label{Eclipse}

Although OpenJML can be used as a command-line tool, much like a compiler, it is also integrated with
an Eclipse plug-in that provides a GUI interface to the type checking, static checking, and runtime assertion checking capabilities of OpenJML. The Eclipse framework is well-supported and mature enough that creating new plug-ins is now a fairly straightforward development task. The author has developed a number of such tools, including interfaces for ESC/Java2 \cite{Kiniry-Cok05-short}, for SMTLIB \cite{Cok-2011-jSMTLIB-short}, and for C/C++ using Eclipse/CDT \cite{Cok-2014-SPEEDY}.
There are a few aspects of the Eclipse GUI for OpenJML that are particularly relevant to the goal of providing an integrated IDE for verification, discussed in the subsections below.

\subsection{Status of verification attempts}
Recall that JML is a modular verification system: for each Java method, the method's implementation is checked against its own specifications and the specifications of referenced types and methods, but without reference to other methods' implementations. It is convenient to launch verification attempts of all the methods in a system, perhaps in parallel, returning later to address individually the methods and classes that show problems. A typical command-line run will show a mix of failed and successful verifications in a long stream of text.

\begin{wrapfigure}[22]{R}{.45\textwidth}
\includegraphics[width=.45\textwidth]{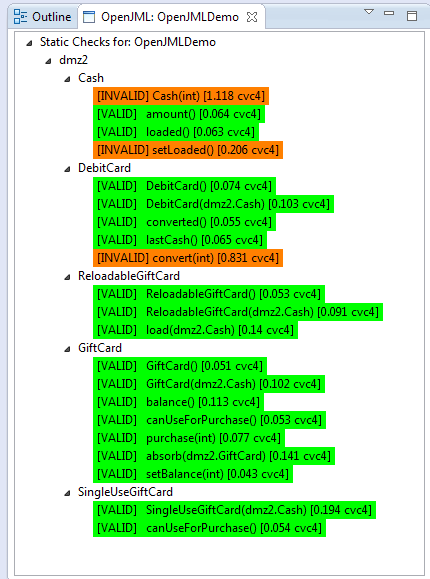}
\caption{Proof Status View, showing the proof attempt results on a collection of methods}
\label{Fig:ProofStatus}
\end{wrapfigure}

Thus the OpenJML GUI provides an Eclipse View that maintains a summary of the results of the most recent verification attempts on each method. When dependency tracking is implemented, this View will also show when a result is out of date. A given method may have multiple falsified assertions; the OpenJML GUI records each
proof attempt for each method. The user can then select among these to see the details (such as
the counterexample values) for any particular proof attempt. A screenshot of the proof status View is shown in Fig. \ref{Fig:ProofStatus}. Each method shows the prover that was used and the time taken in the proof attempt; the color of the tree entry shows the status (valid or invalid or out of date).

\subsection{Showing counterexample paths and values} 
\label{counterexample}
An important innovation in the Eclipse GUI for OpenJML is the ability to explore counterexample information interactively. If a VC is falsified, OpenJML retrieves the counterexample information from
the back-end SMT solver. This information is in terms of the logical variables present in the VC as translated into SMTLIB; those variables and the way the SMT solver expresses its internal values are
generally inscrutable even to experts in the tools involved -- this has been a complaint since ESC/Java \cite{ESCJava2002} used Simplify \cite{Detlefs-Nelson-Saxe03}.

\begin{wrapfigure}[18]{R}{.48\textwidth}
\includegraphics[width=.48\textwidth]{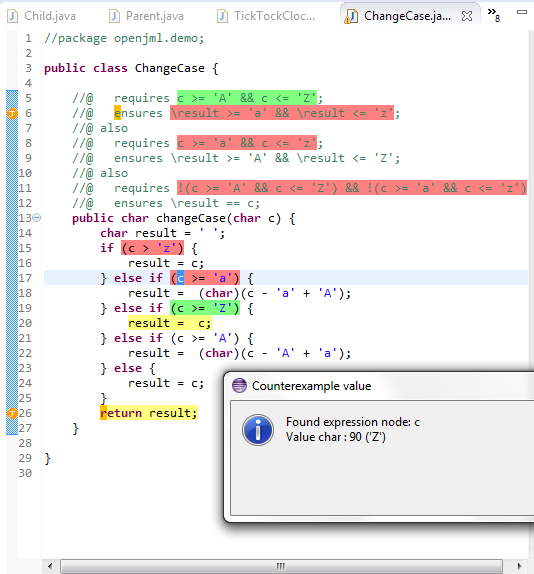}
\caption{Editor window, with highlighted counterexample path and variable value information}
\label{Fig:Counterexample}
\end{wrapfigure}

OpenJML retains information about the mapping of source code expressions into the SMTLIB logical variables. Then the counterexample information, which maps logical variables to values, can be reinterpreted as a mapping of source code expressions to values. Recall that the single-assignment transformation gives different logical variables to different uses of source code variables; thus the values assigned to source variables are the values appropriate to the program state at that point in the program flow. The Eclipse GUI can display this information by means of menu actions or hover information when the mouse is placed over a relevant source code expression. Fig. \ref{Fig:Counterexample} shows a screenshot of the counterexample path for
a buggy method.

In addition, the counterexample contains information about the program path from the start of the method to the assertion failure, since it contains the boolean values for each branch condition. Thus the path to the failed condition in the source code can be determined. OpenJML's GUI displays this information both as a textual program trace (in another Eclipse View) and as highlighting overlaid on the
source code itself in the editor window.

\subsection{Double compilation}
The one disadvantage of this integration with Eclipse is that the Eclipse compiler is executed to show syntax errors and
to compile the .class files, and the OpenJML/OpenJDK compiler is executed to perform the OpenJML actions.
Thus there is double work - the source code is parsed and typechecked twice. The alternative is to
use the Eclipse compiler directly as the front-end for OpenJML. This approach had been tried by both
the author and, independently, by Chalin et al.; both efforts concluded that the Eclipse compiler was
not (at least at that time) suited for extension in the manner needed by JML. On average the extra work is not noticeable compared to the time to execute the proof attempts using the SMT solvers.

\subsection{Other Eclipse functionality}

The Eclipse GUI contains other features that are adapted to specification and verification tasks:
\begin{itemize}[noitemsep,nolistsep]
\item Eclipse problem markers to identify syntactic problems or failed specification checks
\item integration with the Eclipse build system (to automatically perform type checking, static checking, or runtime assertion compilation)
\item menu items to launch various actions, including the ability to map key combinations to actions
\item Eclipse preference and help pages
\item packaging as a standard Eclipse plugin
\end{itemize}
Currently JML type checking is performed when a file is saved; background checking during editing is a planned feature. Static checking is performed either when a file is saved or on manual request. Runtime assertions are compiled, when enabled, along with regular compilation.

\vspace{1em}
\noindent
In the future, OpenJML will acquire some additional GUI features:
\begin{itemize}[noitemsep,nolistsep]
\item syntax highlighting of keywords in specifications;
\item code completion and content assists;
\item integrating specifications into Eclipse's Java refactoring and searching tools.
\end{itemize}

\vspace{1em}
\noindent
A separate project by the author \cite{Cok-2014-SPEEDY} is implementing invariant inference techniques (for C).
OpenJML expects to benefit from this work and to be able to incorporate specification inference as well. That will include features such as these:
\begin{itemize}[noitemsep,nolistsep]
\item automatically suggested specifications;
\item user ability to review, accept, edit, or reject suggested specifications;
\item implementing Eclipse Wizards as a way to guide users through the steps of writing specifications.
\end{itemize}

\section{Building on Software Verification technology}
\label{SVTechnology}
OpenJML intends to be an integration of software verification technology, enabling 
student learning and further research. These research results come from many different researchers, but
are uniquely combined in OpenJML. These various threads include the following:
\begin{itemize}[noitemsep,nolistsep]
\item the Java Modeling Language (JML community) \cite{JMLweb}. Obviously, the Java Modeling Language undergirds OpenJML. Note though that the experience of creating a thorough implementation of JML and
assessing its usability has led to many proposals for changes in JML.
\item Eclipse plug-in framework (adapted from ESC/Java2). The Eclipse GUI for OpenJML has evolved and improved over time from its origins as the GUI for ESC/Java2.
\item basic block/single-assignment translation of programs (Leino and others) and efficient encoding of basic blocks as verification conditions \cite{Leino05,FlanaganSaxe01,Barnett-Leino05-short} . This basic structure was also used in ESC/Java, though the implementation has been streamlined throughout the OpenJML project. The OpenJML project also expects to integrate with Boogie \cite{LGLM:BVD} in the future.
\item non-null types (Chalin, Ernst) \cite{ChalinJR08,chalin2007non,DBLP:conf/issta/PapiACPE08}. JML adopted a default of non-null references.
\item handling of undefined expressions (Chalin) \cite{chalin2005reassessing}. JML originally  assigned arbitrary values to otherwise undefined expressions (such as a division by 0). Chalin's survey of users' interpretations of code and specifications led JML to adopt a different interpretation: 
specifications that cannot be proved to be well-defined are disallowed. So \texttt{i/j < 10} (for unconstrained {\em j}) is undefined, but \texttt{j != 0 \&\& i/j < 10} is well-defined.
\item arithmetic modes (Chalin) \cite{chalin2004jml}. Chalin's proposal of explicitly allowing different arithmetic modes -- pure Java, safe Java (no overflow allowed), and infinite precision arithmetic -- was adopted by JML and is being implemented in OpenJML. JML has syntax allowing different modes to be used in different methods and subexpressions (and should allow different scopes).
\item source code mechanisms for debugging counterexample information (Cok) \cite{Cok-2010-ImprovedUsability}. This innovation is described in section \ref{counterexample}.
\item method calls in specifications (Cok, M\"uller) \cite{Cok04,DarvasM06}. The original ESC/Java inlined called procedures. A scalable system, however, must implement the complex details of modeling called procedures without inlining.
\item datagroups and frame conditions (Leino) \cite{Leino:1998:DGS:286936.286953-short}. JML adopted Leino's work on using datagroups for frame conditions in combination with inheritance and information hiding.
\item observational purity (Cok, Leavens) \cite{Leavens-Cok-05}. JML's rules on purity (absence of side-effects) turn out to be somewhat unworkable in practice, particularly combined with inheritance. Cok and Leavens produced initial proposals for additions to JML to accommodate observational purity.
\item Java generics (Cok) \cite{Cok-2008-SAVCBS}. Cok's initial design for handling generics in JML with Java 5ff is being extended and adapted as it is implemented in OpenJML.
\item vacuity checking (various). A variety of groups used informal techniques to check for specifications causing infeasible paths. OpenJML has implemented the basic technique as a default part of the analysis of a method. 
\item visibility rules (Leavens, M\"uller, Leino) \cite{Leavens-Mueller06}. The interaction of information hiding and specifications is intricate. OpenJML has implemented the rules defined in \cite{Leavens-Mueller06}. This has had a distinct effect on how information hiding and specifications are
presented to students.
\item handling of model fields (Leino, M\"uller) \cite{Leino-Mueller06}. Model fields have values defined by {\em represents} clauses. There are tricky issues regarding when such fields have well-defined values, which can cause unsoundness if implemented poorly.
\end{itemize}
\noindent 
In addition, there are a number of other technologies planned or under consideration for implementation:
\begin{itemize}[noitemsep,nolistsep]
\item Universe types \cite{DietlDM07}
\item JSR 308 \cite{JSR308-webpage-201110}, which allows annotations as part of type identifiers, enabling subtypes to be defined within Java and enabling user-defined type-state-like properties
\item integration with annotation processors
\item multi-threading
\item BML (Bytecode Modeling language) \cite{BML2007-short}, a means to embed specifications in Java byte code 
\item Use and semantics of model programs \cite{ShanerLN07}
\item Java 8
\item Integration with unit testing and Daikon \cite{ErnstPGMPTX2007}
\end{itemize}

\section{Status of the implementation of OpenJML}

As mentioned already, OpenJML is still a work in progress; nearly all of the development has been an
unfunded volunteer effort. Nevertheless, the project has proceeded to the point to be useful in educational settings and for small scale case studies on existing code. Not all features of Java and JML are yet implemented. This section gives a summary of the status as of the submission of this paper. Updates can be found on the OpenJML web site (\url{www.openjml.org}). In general, all Java and JML language elements are parsed and type-checked, but those not implemented are ignored for static or runtime checking.

\subsection{Java Language elements}

\noindent \textbf{Java 4:} All language elements are implemented except
\begin{itemize}[noitemsep,nolistsep]
\item features of multi-threaded Java
\item the strictfp modifier (and floating-point semantics in general)
\item the volatile, transient, and native modifiers (and details of memory models)
\end{itemize}

\noindent \textbf{Java 5:} These are the new features in Java 5:
\begin{itemize}[noitemsep,nolistsep]
\item Generic types - partially implemented
\item enhanced for loop - implemented
\item autoboxing and unboxing - implemented
\item typesafe enums - simple enum classes are implemented
\item static import - implemented (but JML model imports are treated just like Java imports) 
\item varargs - partially modeled 
\item Java annotations -  No checks on annotations are modeled by ESC or compiled by RAC
\end{itemize}

\noindent \textbf{Java 6:} No language changes

\noindent \textbf{Java 7:} New Java 7 language elements:
\begin{itemize}[noitemsep,nolistsep]
\item binary literals - implemented
\item literals with underscores - implemented
\item Strings in switch statements -  Compiled in RAC, but not modeled or checked by ESC
\item try with resources - Compiled in RAC, but not modeled or checked by ESC 
\item Catching multiple exception types - Compiled in RAC, but not modeled or checked by ESC
\item Type inference - not implemented in JML
\item Runtime errors associated with varargs parameter lists - not implemented in ESC
\end{itemize}

\noindent \textbf{Java 8:} (OpenJML is only implemented at present for Java 7). These new language features will be required to migrate to Java 8:
\begin{itemize}[noitemsep,nolistsep]
\item lambda expressions (closures)
\item JSR308 and other enhancements to Java annotations
\end{itemize}

\subsection{JML language elements}
The Java Modeling Language has a large number of features (over 100 groups of features). To list the status of all of them here would be space-consuming, quite quickly out of date, and not
particularly interesting in the details. The reader is referred instead to the OpenJML web page
(\url{www.openjml.org}) and the specific page listing implementation status for current status at future times.

The most important missing features are these:
\begin{itemize}[noitemsep,nolistsep]
\item Basic (level 0) JML is nearly all implemented, with a few restrictions. The main gaps are
the implementation of the Universe type system and encoding and SMT solver support for {\tt {\textbackslash}sum}, {\tt {\textbackslash}product}, and {\tt {\textbackslash}num\_of} quantifiers. The maps clause (level 1) is needed for more substantial examples. Checks on static and instance initialization are still in progress.

\item There is significant desire to include the proposed {\tt {\textbackslash}past} operator in standard JML and OpenJML, as an improvement on the traditional {\tt {\textbackslash}old}.

\item Key advanced features that are waiting for implementation include model programs, some of the less commonly used method specification clause types, and object set predicates such as {\tt {\textbackslash}reach}.

\item Support for concurrency still requires research, experimentation and trial use before it is 
ready to be implemented.

\item Support for specifications in class files (as in BML \cite{BML-2009}) has not yet been investigated.

\end{itemize}

\vspace{1em}
Despite the missing features, the current status is sufficient to be applied to interesting educational examples and small-scale use cases of industrial code.

\subsection{Soundness}

OpenJML intends to be a sound tool. That is, it intends to correctly encode the semantics of Java and JML. However, the soundness and completeness of all software verification tools is limited in various ways. OpenJML has addressed some of the sources of unsoundness and incompleteness that were present in ESC/Java (cf. Appendix C in \cite{Leino-Nelson-Saxe00}, from which some of the following discussion is derived). A user of OpenJML experimenting with Java and JML will be concerned about two questions: (a) does every inconsistency between the software and the specifications trigger a warning by OpenJML; (b) does every warning by OpenJML correspond to a fault in the code+specifications.\footnote{These two questions correspond to the questions of soundness and of completeness, but which label is applied to which question varies depending on perspective.}

\paragraph{Does every inconsistency between the software and the specifications trigger a warning by OpenJML, that is, are there missed errors (false negatives)?}
\begin{itemize}[noitemsep,nolistsep]
\item Note that one cause of missed errors is incorrect specifications. Wrong specifications can introduce
spurious errors, but they can also make program paths infeasible, and thereby hide errors. That is 
why OpenJML checks for feasibility when the first check of a method indicates no inconsistencies between specifications and code.
\item Furthermore, OpenJML verifies each method independently, presuming that all the specifications it uses, including those of other methods, are correct. An error in method A's spec may cause errors to be 
hidden in method B. All specifications and methods must be consistent simultaneously.
\item The semantics of JML state that all invariants of all classes and objects must hold at the 
beginning of any method. In practice, OpenJML assumes all the invariants of a heuristically-determined relevant set of classes. This has the potential of missing a relevant invariant that might constrain
the logical state enough to uncover an error. This source of false negatives is less prevalent if invariants are `well-behaved', that is, they refer only to fields of their own objects. Implementation of the ownership type system will also mitigate this problem.

\item OpenJML relies on SMT solvers to find counterexamples in a collection of logical assertions that
represent the program and specifications. The solver may exhaust available time or memory resources; 
in that case, it can at most state that the solver does not know whether there is a specification 
inconsistency or not. This is particularly the case when there are quantified assertions, in which case the logical validity problem may be undecidable. Also, of course, the solver may have bugs.
\item Finally, of course, OpenJML is not complete, and the portions that are complete may have bugs.
\end{itemize}

\paragraph{Does every warning by OpenJML correspond to a fault in the code+specifications, that is, are there spurious warnings (false positives)?}

\begin{itemize}[noitemsep,nolistsep]
\item There is an intrinsic source of false positives, namely the incompleteness of the underlying logical theories. This allows the underlying solvers to find `counterexamples' that are not valid in reality. This is particularly the case when quantified expressions are used. 

\item The above is exacerbated by the fact that some of Java's behaviors have yet to be modeled by JML or the underlying SMT provers (floating point semantics are one example).
\item Finally, the modular nature of JML's modeling may preclude some deductions that might be possible with a different style of specification and specification checking.
\end{itemize} 

\section{Uses and Experience}

To date, there have been these principal uses of OpenJML:
\begin{itemize}[noitemsep,nolistsep]
\item OpenJML was used as the basis for student assignments in software verification in at least four different courses during the fall of 2013, with others planned in 2014. These courses were undergraduate courses that used formal methods or formal specifications (cf. Section \ref{Acknowledgments}).
\item OpenJML is used as a basis for research on information flow on the FlowSpecs project (Naumann and Leavens), funded by NSF SaTC grants CNS1228930 and CNS1228695 (\url{http://www.cs.stevens.edu/~naumann/xxxflowspecs/index.html}).
\item OpenJML is integrated into JmlUnitNG \cite{Zimmerman2010}
\item Use cases of specification and verification of medium-scale Java software are in progress.
\item The API has been used (in projects of Kiniry's students) to embed OpenJML in other systems.
\end{itemize}

\begin{wrapfigure}[31]{r}{.5\textwidth}
\small
\begin{verbatim}public class A {
  private int _value;
  
  //@ assignable this.*; // default
  public A(int v) {
    _value = v;
  }
}
           (a)
           
public class A {
  private int _value; //@ in value;
  //@ model public int value; 
  //@ public represents value = _value;
  
  //@ assignable this.*; // default
  public A(int v) {
    _value = v;
  }
}
            (b)
            
public class A {
  //@ public model Object \state;//implicit
  private int _value;
  //@ in \state; // implicit
  
  //@ assignable this.*; //includes \state
  public A(int v) {
    _value = v;
  }
}
             (c)
\end{verbatim}
\caption{(a) Example default assignable clause; (b) with model field; (c) with default model field}
\label{Fig:DefaultAssignable}
\end{wrapfigure}

\vspace{1em}
Informal experience with OpenJML so far is that, first, the GUI is a major help in editing, checking, and debugging specifications. Second, implementing verification technology for a production language, such as Java,
while perhaps more complex for the implementors and users, holds the promise of applying verification technology in actual software development. And third, the integration of the variety of technologies listed in section \ref{SVTechnology} is a useful (though lengthy) exercise. The exercise of implementing the combination exposes ambiguities in understanding and definition of features. It also exposes complexities in their interactions. One of those is presented in the following subsection.

\subsection{Frame condition defaults for constructors}

Part of the specification of a method is a statement of the frame condition (the {\em assignable} clause), which states what memory locations (variables, field elements, array elements) the method may modify. For non-constructor methods not marked as pure, the default (when no assignable clause is given) is that the method may modify anything ({\tt assignable {\textbackslash}everything;}). For constructors, the default defined by JML is {\tt assignable this.*;}. 
Originally, {\tt this.*} was defined to mean a list of all the (non-static) fields of the object. This was a
useful default, because a typical constructor initializes all the fields of the object.
However, this behavior interacts with JML's rules about visibility. One visibility rule is that a
specification clause may not refer to variables of more restricted visibility than the clause itself; a client that can see the clause must be able to see the variables contained in the clause. Thus, a public 
assignable clause may refer only to public fields of the object. So, in a public
{\tt assignable this.*;} clause, the {\tt this.*} was redefined to expand only to the public fields of the object.
As a result, straightforward code such as in Fig. \ref{Fig:DefaultAssignable}(a), will result in a
warning: the constructor modifies a field that it is not permitted to. A student might very well write such code as a first attempt to use JML.

There are two remedies for this situation, both of which add complexity for the beginning specifier.
The private {\tt \_value} field can be declared {\tt spec\_public}. Alternately, and equivalently,
we can declare a model field serving as a datagroup, and place the private field {\em in} the datagroup,
as in Fig. \ref{Fig:DefaultAssignable}(b). Both of these solutions require some understanding of the interaction of private implementations with public specifications.

%
%

A third alternative is to modify JML by adding an implicit model field to every object. The model field, as a datagroup,
would contain all the fields of the object (perhaps JML should define {\tt this.*} to be just that model field). This is the equivalent of having
the declarations in Fig. \ref{Fig:DefaultAssignable}(c) be present by default in every object's specification. Such a solution would permit the original example 
to
pass without modification; however, it simply hides complexity behind defaults chosen to improve usability.

%

\section{Related work}

OpenJML builds on ESC/Java \cite{ESCJava2002} and ESC/Java2 \cite{Kiniry-Cok05-short}.
Some static analysis capabilities are now being built into compilers such as Eclipse and Visual Studio.
However, the only tools that provide automated checking of functional specifications are Spec\# \cite{Barnett-Leino-Schulte04-short} for an extension of C\#, Frama-C \cite{webframac} for C with ACSL \cite{ACSLweb} specifications, tools for SPARK/Ada and, more recently, Dafny \cite{Leino:Dafny:LPAR16-short,Dafny2014}. Eiffel \cite{Meyer88} built less expressive specifications directly into the programming language. Leon \cite{Blanc:2013:OLV:2489837.2489838-short} is a verifier for Scala programs. Key \cite{KeYBook2007} and Why3 \cite{Why3} are other recent tools that apply to Java programs, but require the user to interact with theorem proving tools separately from the software development IDE. The Mobius project \cite{webmobius} combined a large number of tools (including ESC/Java2) into a suite
of verification tools.

\section{Future work}

The work in the immediate future falls into four areas:
\begin{itemize}[noitemsep,nolistsep]
\item Completing the modeling of the features of basic JML sequential Java, and concurrent Java
\item Migrating to Java 8 and supporting its significant new features (closures and annotations)
\item Adding some additional tools, valuable for both education and specification development, such as automatic test generation and documentation
\item Enhancing documentation, especially tutorials and instructional material
\end{itemize}

\vspace{1em}
But the more interesting activities are what OpenJML enables. First is work on invariant inference. One goal of an industrial-strength verification system must be to minimize the manual effort and cognitive load on the user by automating as many tasks as possible. Integrating tools to generate candidate specifications from the source code (or, Daikon-like, from test runs), will help to lessen the work
currently needed to successfully specify a substantial software application.

A second important area is the thoughtful execution of use cases. The amount of research on verification techniques far exceeds the amount of application of verification. Verification challenges help elucidate the methods needed to handle common programming idioms. In addition, however, tackling real verification problems (in all their messiness) is 
essential to discovering the techniques, default behaviors, and UI mechanisms that are needed in practice.

A third effort needs to be education in software verification. Adoption of verification techniques as
part of daily software development is slow - in part because today's tools are inadequate, but also
because of unfamiliarity on the part of software developers. More and better tutorials, education materials, undergraduate and graduate courses, and expectations for actual practice can progress the
goal of more application of verification technology and hence better software.

\section{Acknowledgments}
\label{Acknowledgments}

OpenJML is developed by David Cok, with encouragement from the JML community and others.
\begin{itemize}[noitemsep,nolistsep,leftmargin=*]
\item Dan Zimmerman (Harvey-Mudd College), Wojciech Mostowski (University of Twente), Marieke Huisman (University of Twente), and others provided valuable usability feedback and bug reports as part of courses they have taught that used OpenJML. 
\item Student projects by Alysson Milanez and Tiago Massoni contributed test cases.
\item An equipment grant from aicas GmbH (\url{www.aicas.com}) enabled MacOS development. 
\item A small amount of development was performed under NSF SaTC grants, including work by John Singleton on configuration checking: CNS1228930 and CNS1228695. The paper is also based partially on work supported by the National Science Foundation under Grant No. ACI-1314674. Any opinions, findings, and conclusions or recommendations expressed in this material are those of the author(s) and do not necessarily reflect the views of the National Science Foundation.
\item OpenJML
is an implementation of JML (\url{www.jmlspecs.org}), makes use of SMTLIB (\url{www.smtlib.org}) and is built on both OpenJDK (\url{openjdk.java.net}) and Eclipse  (\url{www.eclipse.org}). OpenJML also makes use of SMT solvers
from various groups, most particularly, CVC4, Z3, and Yices.

\end{itemize}

\bibliographystyle{eptcs}

\end{document}